\documentclass[aps, prl,showpacs,amsmath,amssymb,amsfonts,twocolumn,superscriptaddress,floatfix,footinbib]{revtex4-1}

\usepackage{amsmath}
\usepackage{amssymb}
\usepackage[final]{graphicx}
\usepackage{upgreek}
\usepackage[utf8]{inputenc}

\newcommand{\mueV}{\,\upmu\mathrm{eV}}
\newcommand{\mum}{\,\upmu\mathrm{m}}

\newcommand{\distance}{5mm}			
\usepackage{color}

\begin{document}
\title{Phonon--assisted Lasing in ZnO Microwires at Room Temperature}
\author{T. Michalsky}
\affiliation{Universität Leipzig, Institut für Experimentelle Physik II, Linnéstraße 5, 04103 Leipzig, Germany}
\author{M. Wille}
\affiliation{Universität Leipzig, Institut für Experimentelle Physik II, Linnéstraße 5, 04103 Leipzig, Germany}
\author{C. P. Dietrich}
\thanks{Present address: University of St Andrews, School of Physics and Astronomy, North Haugh, KY16 9SS, St Andrews, Fife, Scotland, United Kingdom}
\affiliation{Universität Leipzig, Institut für Experimentelle Physik II, Linnéstraße 5, 04103 Leipzig, Germany}
\author{R. Röder}
\affiliation{University of Jena, Institute for Solid State Physics, Max-Wien-Platz 1, 07743 Jena, Germany}
\author{C. Ronning}
\affiliation{University of Jena, Institute for Solid State Physics, Max-Wien-Platz 1, 07743 Jena, Germany}
\author{R. Schmidt-Grund}
\affiliation{Universität Leipzig, Institut für Experimentelle Physik II, Linnéstraße 5, 04103 Leipzig, Germany}
\author{M. Grundmann}
\affiliation{Universität Leipzig, Institut für Experimentelle Physik II, Linnéstraße 5, 04103 Leipzig, Germany}
\date{\today}

\begin{abstract}
\noindent We report on room temperature phonon--assisted whispering gallery mode (WGM) lasing in ZnO microwires. For WGM laser action on the basis of the low gain phonon scattering process high quality resonators with sharp corners and smooth facets are prerequisite. Above the excitation threshold power~\textit{P}$_{\text{\textit{Th}}}$ of typically 100\,kW/cm$^2$, the recombination of free excitons under emission of two longitudinal optical phonons provides sufficient gain to overcome all losses in the microresonator and to result in laser oscillation. This threshold behavior is accompanied by a distinct change of the far and near field emission patterns, revealing the WGM related nature of the lasing modes. The spectral evolution as well as the characteristic behavior of the integrated photoluminescence intensity versus the excitation power unambiguously prove laser operation. Polarization-resolved measurements show that the laser emission is linear polarized perpendicular to the microwire axis~(TE). 
\end{abstract}


\maketitle
\noindent\hspace*{\distance}Extensive research on the wide band gap semiconductor zinc oxide (ZnO) started in the 1950s. Hence, this material system is well explored \cite{Morkoc2008,Jagadish2011} but is still in the focus of current research \cite{Wang2012,Piccione2014}. Already in 1973 Klingshirn reported on the different laser processes in bulk ZnO and their temperature dependence~\cite{Klingshirn1973}. Above~120~K, the main extent of bound excitons is dissociated and the thermal occupation of longitudinal optical~(LO) phonon states leads to an enhanced absorption in the energy range between the free exciton energy and its first \mbox{LO--phonon} replica \mbox{E$_{\text{FX}}$\,--\,E$_{\text{FX--1LO}}$}. Hence, a transition from \mbox{FX--1LO} to \mbox{FX--2LO} lasing was observed at temperatures above~120~K. The respective \mbox{FX--2LO} recombination process was found to be stable up to 250~K in macroscopic single crystals~\cite{Klingshirn1973, Nicoll1966, Iwai1970}. At room temperature, the 2LO--phonon--assisted recombination of free excitons exhibits relatively low gain values with $\alpha \sim 50~\text{cm}^{-1}$~\cite{Klingshirn1975} in contrast to the other possible gain processes such as exciton--exciton scattering with~$\alpha \sim 300~\text{cm}^{-1}$ \cite{Tang1998} and electron--hole plasma (EHP) with $\alpha \sim 10^3-10^4~\text{cm}^{-1}$~\cite{Bohnert1980}. Thus, competing processes like exciton--exciton or exciton--electron scattering already provide higher gain values at low excitation densities~\cite{Klingshirn1973}. The \mbox{FX--2LO} laser process and its temperature dependence remained difficult to access experimentally, in particular at room temperature.\\
\noindent\hspace*{\distance}Of major interest in the last few years’ research was the investigation of stimulated emission in nano- and microstructures with the goal to achieve 1D and 2D light emitting devices and laser diodes in the near UV spectral range~\mbox{\cite{Chu2011, Huang2001, Czekalla2008, Dai2009}}. In fact semiconducting ZnO microwires~(MWs) provide all components necessary for laser action: Under sufficiently high excitation the semiconducting material provides gain and the hexagonal cross section of the MWs represents a whispering gallery type resonator due to total internal reflections (TIR) at the inner sidewalls. The propagation of whispering gallery modes (WGMs) in hexagonal ZnO MWs was demonstrated in Ref.~\cite{Czekalla2008, Nobis2004}. 
\begin{figure}[t]
\centering
\includegraphics[width=8.4cm]{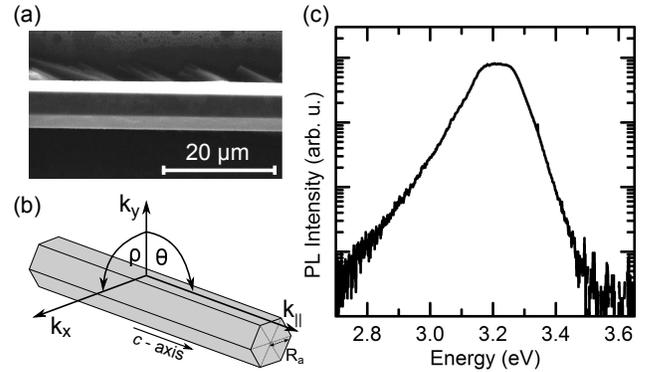}
\caption{(a) Scanning electron microscope image of a typical ZnO microwire, which is lying on a sapphire substrate. This is the wire from which the optical observations are presented. (b) Sketch of a hexagonal wire including angle and wave vector definitions used in this paper. (c) Typical room temperature PL spectrum for $k_{\parallel}=0$ at low excitation densities.}
\label{wire_spek}
\end{figure}
As TIR exhibit no losses, all optical losses are connected to the wire corners and to surface irregularities~\cite{Wiersig2003}. This leads to high quality factors ($Q=\lambda/\Delta\lambda$) and low lasing thresholds in MWs~\cite{Dietrich2012}.\\ 
\noindent\hspace*{\distance}In this work, we demonstrate room temperature \mbox{FX--2LO} lasing in a ZnO MW resonator. The resonator quality in terms of morphological properties of the MW is crucial to realize \mbox{2LO--phonon}--assisted laser operation. The impact on the WGM occupation as well as on the emission characteristics in~\mbox{\textit{k}-space} and real-space, after crossing the laser threshold, will be discussed in detail.\\
\noindent\hspace*{\distance}The ZnO MWs
\begin{figure*}[hbtp]
\centering
\includegraphics[width=15cm]{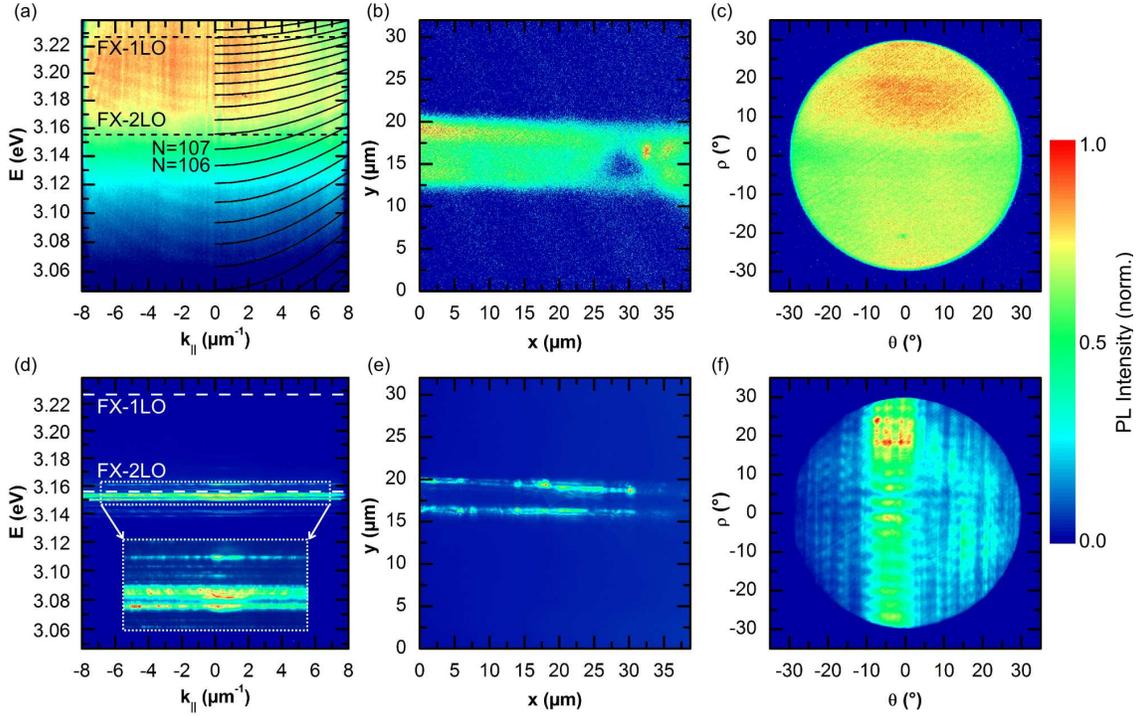}
\caption{Optical emission of a ZnO microwire below (0.2\textit{P}$_{\text{\textit{Th}}}$: a,b,c) and above (2\textit{P}$_{\text{\textit{Th}}}$: d,e,f) threshold power~\textit{P}$_{\text{\textit{Th}}}\sim$~90\,kW/cm$^2$ at room temperature. Overcoming \textit{P}$_{\text{\textit{Th}}}$ changes the energy resolved \mbox{k-space} characteristic from (a)~weak WGM emission in the entire energy range to (d)~high intensity, narrow and dispersionless emission around~3.15\,eV. The homogeneous emission over the wire in (b) changes to emission only from the corners in (e) and the isotropic angular emission pattern in (c) converts to interference pattern in (f). The edge emission causes the obtained interferences in $\rho$ direction similar to a “Young double slit”.}
\label{Bild1}
\end{figure*}
were grown by carbothermal evaporation of a pressed target consisting of zinc oxide and carbon in a mass ratio of~1:1. The target pellet was put in a tube furnace, heated up to 1400\,K with a temperature ramp of 300\,K/h. Growth took place for~1\,h and the furnace was cooled down afterwards with -300\,K/h. The entire growth process took place in ambient air. The MWs with diameters in the \mbox{$\mum$ range} and lengths up to several~mm grew directly on the target pellet. Subsequently, the MWs have been transferred and fixed on a sapphire substrate. This allows for the successive investigation of morphological and optical properties of single MWs. Figure~\ref{wire_spek}(a) shows a scanning electron microscopy (SEM) image of a typical ZnO MW. The photoluminescence spectroscopy was performed using the fourth harmonic of a Nd:YAG laser (266\,nm, 20\,Hz repetition rate, 10\,ns pulse duration) focused to a spot with a diameter of around~150$\mum$ on the sample. The excitation spot is much larger than the diameter of the investigated MWs, therefore the excitation can be regarded as homogeneous over the investigated area. A variable attenuator was used to regulate the excitation power density, while the wires were excited under an angle of incidence of~$\sim$\,60\,$^\circ$ to the substrate normal. The luminescence from the sample was collected with a wavelength-corrected~NUV objective (NA\,=\,0.5; detectable angular range of $\pm$\,30\,$^\circ$), dispersed by a spectrometer (320~mm focal length, 2400\,groves/mm grating), and detected by a Peltier-cooled, back-illuminated CCD camera. A variable lens setup allowed imaging of both, the real and \mbox{\textit{k}-space}.  This setup had a spectral resolution of~500$\mueV$ and an angular resolution of about~0.5\,$^\circ$.\\
\noindent\hspace*{\distance}All results presented below were obtained from the ZnO~MW shown in Fig.~\ref{wire_spek}(a) with an outer radius of~\mbox{$R_{out}\sim 3.85\mum$}. This wire exhibits sharp corners and a smooth surface, hence extraordinary morphological properties in terms of a high quality optical WGM resonator~\cite{Dietrich2012}. 
Under low excitation density the main emission of the wire is located in the energy range of \mbox{3.1\,--\,3.3\,eV}, as can be seen in the PL spectrum for  $k_{\parallel}=0$ in Fig.~\ref{wire_spek}(c). Energy resolved~\mbox{$k_{\parallel}$-space} imaging reveals WGMs propagating in the microresonator, as can be seen in~Fig.~\ref{Bild1}(a). Using the plane wave model~\cite{Wiersig2003a} and a parabolic dispersion relation 
\mbox{$E(k_{\parallel})= \hbar c\sqrt{k_{N}^2+\left(k_{\parallel}\slash n_{\perp\slash\parallel} \right)^2}$}, the WGM can be assigned to mode orders~\textit{N} in the range of \mbox{100\,--\,118}. Here, $k_{\parallel}=\frac{2\pi}{\lambda}\sin{\theta}$ is the wave vector component parallel to the wire axis (compare with~Fig.~\ref{wire_spek}(b)), $k_{N}$ denotes the resonant wavenumbers with interference order~\textit{N}, $n_{\perp\slash\parallel}$~is the polarization dependent refractive index \footnote{It should be noted that the refractive index which was used to calculate the mode numbers was obtained by spectroscopic ellipsometry on thin film ZnO grown by pulsed laser deposition \cite{Schmidt-Grund2011}, thus it might differ from the refractive index of the wire, possibly causing some uncertainty in the obtained radius and mode numbers.} and $\perp$/$\parallel$~denotes polarization perpendicular/parallel to the \textit{c}--axis of the wire, respectively. Furthermore, under low excitation densities the MW emission is spatially homogeneous over the whole wire surface and isotropic in all detectable emission angles (Figs.~\ref{Bild1}(b) and~(c)). If the excitation power is increased above a certain value~\textit{P}$_{\text{\textit{Th}}}$, the situation changes dramatically. The main emission is now strongly localized within a few, very narrow emission lines extending over the full observable $k_{\parallel}$-range around the energy~3.15\,eV, as shown in~Fig.~\ref{Bild1}(d). Furthermore, within these dispersionless lines the intensity is highest when its \textit{k}-value matches the dispersion relation of a WGM. The spatial emission above threshold is dominated by the emission out of the wire corners (see Fig.~\ref{Bild1}(e). The angular-resolved emission in Fig.~\ref{Bild1}(f) clearly exhibits an alternating intensity pattern, in contrast to the isotropic emission below threshold. Furthermore polarization measurements above~\textit{P}$_{\text{\textit{Th}}}$ reveal, that the integrated PL~signal is mainly polarized perpendicular to the MW \textit{c}--axis, with a polarization ratio of~\mbox{$P=(I_{\perp}-I_{\parallel})/(I_{\perp}+I_{\parallel})=0.78$}, which is about a factor of 3 larger than below~\textit{P}$_{\text{\textit{Th}}}$.\\ 
\noindent\hspace*{\distance}The emission behavior below threshold can be explained by the spontaneous decay of excitons having no restrictions for the emission direction. WGMs are fed by this spontaneous decay and can propagate inside the wire cross section. Their interaction with the excitons is reflected in the flattening of the dispersion relations and narrowing of the mode spacings with increasing mode number~\textit{N}~\cite{Dietrich2011}. Crossing the excitation threshold, the main emission energy of~3.15\,eV coincides with the energetic position of the free A and B excitons ($E_{FX_A}$\,$\sim$\,3.300\,eV; $E_{FX_B}$\,$\sim$\,3.305\,eV~\cite{meyer2004bound}) under emission of two~\mbox{LO--phonons} ($E_{LO}$\,$\sim$\,72\,meV). This process is attributed to cause gain in a dispersionless spectral band near~3.15\,eV since the exciton and phonon dispersions in our observable $k$-range are here almost flat~\cite{meyer2004bound, Serrano10}. The laser action starts for WGMs being in resonance with the corresponding spectral gain region, if the exciton density is high enough that the resulting gain can overcome all resonator losses. The fact that light above~\textit{P}$_{\text{\textit{Th}}}$ is predominantly emitted at the MW corners is in agreement with theoretical predictions~\cite{Wiersig2003} for the outcoupling of WGMs. It is very interesting that the emission stemming from the MW corners interferes, similarly to a double slit with a slit distance of 3.85$\mum$. This leads to a modulated far field emission pattern, as visible in Fig.~\ref{Bild1}(f). Finally, the perpendicular polarization of the emitted light with respect to the wire axis below threshold can be explained by the spontaneous recombination of A and$\slash$or B excitons as they emit photons mainly with this polarization~\cite{Yoshikawa1997}. In the stimulated emission regime the PL~spectra are dominated by WGMs which are linear polarized perpendicular to the wire \textit{c}--axis. Hence the polarization ratio is significantly enhanced in the stimulated emission regime. The fact that the polarization ratio $P=0.78$ is smaller than 1 can be explained by the underlying spontaneous emission which partly randomizes the polarization. If the lasing emission can be attributed to just one of the two excitons (A or B), cannot be clarified in the frame of this work, but probably only the A exciton is involved since it is lowest in energy and is therefore more likely occupied in PL experiments.\\ 
\noindent\hspace*{\distance}The threshold characteristic yield as function of the excitation power density varying in a range from~\mbox{1\,--\,500\,kW/cm$^2$} clearly demonstrates the onset of laser action in the microresonator. 
\begin{figure}[t]
\centering
\includegraphics[width=8.4cm]{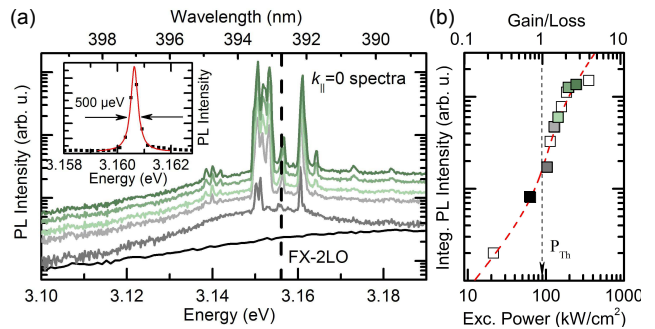}
\caption{(a) Excitation dependent spectra at room temperature for $k_{\parallel}=0$. With increasing excitation power sharp modes appear around the position of the \mbox{FX--2LO} (dashed line) and dominate the overall spectrum. (b) The log-log~plot of the integrated PL intensity vs. the excitation power exhibits a distinct laser characteristic. The dashed red line corresponds to the adapted multimode laser model~\citep{Casperson1975} with a threshold of around 90\,kW/cm$^2$.}
\label{Lasingspektren}
\end{figure}
The spectra in~Fig.~\ref{Lasingspektren}(a) depict the transition from a broad spontaneous emission to stimulated emission in an energy range around the \mbox{FX--2LO} emission energy without any noticeable energy shift of the spectral gain distribution. Note the logarithmic scale in~Fig.~\ref{Lasingspektren}(a) and thus the intensity difference of two orders of magnitude between stimulated mode emission and the underlying spontaneous background at high excitation power. The spectral width of the appearing lasing modes of~\mbox{FWHM\,=\,500$\mueV$} is at the resolution limit of our setup and indicates a high resonator quality~($Q\geq6300$). The double-logarithmic plot of the integrated PL intensity versus the excitation power in~Fig.~\ref{Lasingspektren}(b) depicts a distinct \mbox{S--shaped} course, caused by the threshold behavior of the underlying laser process and allows the determination of the excitation threshold power~\textit{P}$_{\text{\textit{Th}}}\sim$~90\,kW/cm$^2$. Dai et al.~\cite{Dai2009} published room temperature WGM lasing in similar ZnO MWs with a lasing threshold of~\textit{P}$_{\text{\textit{Th}}}$\,=\,255\,kW/cm$^2$ and a FWHM of the lasing modes of~2.4\,meV~($Q\sim 1300$). This highlights the high quality of our sample.\\
\noindent\hspace*{\distance}In literature there are three more processes under discussion, which are able to generate enough gain to overcome the overall losses in ZnO cavities at room temperature. The first highly discussed process is the formation of an electron--hole plasma, which leads to a renormalization of the band gap and thus to a spectral red--shift of the main emission with increasing excitation power~\cite{Lysenko1975}. This could not be observed in our experiments. Furthermore, if each impinging photon creates an electron hole pair and neglecting non--radiative recombination processes, the maximal carrier density at an excitation power of~100\,kW/cm$^2$ can be estimated to be $\sim 10^{18}\,\text{cm}^{-3}$. This upper limit is slightly smaller compared to the Mott density of ZnO given in the literature~\cite{Chen2001, Johnson2003, Klingshirn2010}. Hence we can exclude the EHP to be the underlying gain process in our experiments. The second process is scattering between excitons which leads to an emission band in the energy range of~\mbox{3.28\,--\,3.34\,eV}, which does not fit our observations. If we consider that the 7.7$\mum$ thick MW is pumped only in the outermost region (50\,nm below the surface), there is a huge low--excited volume that does not contribute to the optical gain. Furthermore the reabsorption in this volume is significantly enhanced in the energy range close to the free exciton emission. Hence the laser condition cannot be fulfilled for the exciton--exciton scattering process due to high absorption losses. In the energy range of the FX--2LO emission the absorption in the low--excited volume is almost negligible. Hence the FX--2LO process is much more favorable. The scattering of free electrons with excitons is an important mechanism in the temperature range of \mbox{70\,--\,150\,K}~\cite{Klingshirn1973}, but the low achievable gain value at room temperature as well as the totally different energetic position exclude this process in our experiments. The low lasing threshold of~\textit{P}$_{\text{\textit{Th}}}\sim$~90\,kW/cm$^2$, the energetic position and the non existing renormalization effects confirm \mbox{FX--2LO}~scattering to be the underlying gain mechanism of the observed laser process.\\
\noindent\hspace*{\distance}In summary, we presented 2LO--phonon--assisted WGM lasing at room temperature in a ZnO MW revealing extremely high resonator quality. The mode occupation, the~\mbox{\textit{k}-space} and the real space emission characteristics below and above the laser threshold have been shown. Furthermore, we confirmed the laser action by the excitation dependent spectral evolution and argued that the underlying gain mechanism is the \mbox{FX--2LO} process.\\
\noindent\hspace*{\distance}This work was supported by the Deutsche Forschungsgemeinschaft within GR 1011/26--1, the "Leipzig Graduate School of Natural Sciences~--~BuildMoNa" and through FOR1616. We thank H. Franke, C. Sturm and M. Dorn for valuable discussions and G.~Ramm for the preparation of the ZnO/C pellets.


%

\end{document}